\documentclass{WileyMSP-template} 
\usepackage{amsmath}
\usepackage{xcolor}
\usepackage{booktabs}
 \usepackage{multirow} 
\begin{document}

\pagestyle{fancy}
%\rhead{\includegraphics[width=2.5cm]{vch-logo.png}}

\title{Transfer Learning for Transformer-Based Modeling of Nonlinear Pulse Evolution in Er-Doped Fiber Amplifiers}

\maketitle

% Author: Please give full first and last names for authors and include * after the name of all corresponding authors

\author{A. E. Bednyakova*}
\author{A. S. Gemuzov}
\author{M. S. Mishevsky}
\author{K. P. Saraeva}
\author{A. A. Redyuk}
\author{A. A. Mkrtchyan}
\author{A. G. Nasibulin}
\author{Y. G. Gladush}

% Dedication

%\dedication{Optional dedication here. If no dedication is required, please leave blank}

% Affiliations: Please provide adacemic titles (Prof. or Dr.) for all authors where applicable, and include an institutional email address for all corresponding authors
\begin{affiliations}
A. E. Bednyakova, A. S. Gemuzov, K. P. Saraeva, A. A. Redyuk\\
Novosibirsk State University
Novosibirsk 630090, Russia\\
Email Address: a.bedniakova7@g.nsu.ru

M. S. Mishevsky, A. A. Mkrtchyan, A. G. Nasibulin, Y. G. Gladush\\
Skolkovo Institute of Science and Technology\\
Skolkovo 121205, Russia
\end{affiliations}

% Keywords: Please provide a minimum of three and a maximum of seven keywords, separated by commas


% Abstract should be written in the present tense and impersonal style (i.e., avoid we), and be at most 200 words long
\begin{abstract}
A neural network model based on the Transformer architecture has been developed to predict the nonlinear evolution of optical pulses in Er-doped fiber amplifier under conditions of limited experimental data. To address data scarcity, a two-stage training strategy is employed. In the first stage, the model is pretrained on a synthetic dataset generated through numerical simulations of the amplifier’s nonlinear dynamics. In the second stage, the model is fine-tuned using a small set of experimental measurements. This approach enables accurate reproduction of the fine spectral structure of optical pulses observed in experiments across various nonlinear evolution regimes, including the development of modulational instability and the propagation of high-order solitons.
\end{abstract}

% Text: Please use section headings and subheadings as specified below. For communications, all section headings apart from Experimental Section should be removed
% Please make the first reference to a display item bold: \textbf{Figure 1}
% Do not abbreviate Figure, Equation, etc.; display items are always singular, i.e., Figure 1 and 2.
% Equations are always singular, i.e., Equation 1 and 2, and should be inserted using the {equation} environment, not as graphics
% Please do not use footnotes in the text, additional information can be added to the Reference list.

\section{Introduction}

The nonlinear evolution of ultrashort pulses with high peak power in fiber laser systems is of great practical importance, as these pulses are widely used across various scientific and technological fields \cite{richardson2010, fu2018}. Applications include material processing, such as material structuring and surface texturing (for controlled wettability, friction, etc.), precision cutting (e.g., semiconductor chips, transparent materials like glass and ITO), microdrilling (microfluidic devices, injector nozzles) \cite{ORAZI2021543}. Another type of application employs ultrafast lasers for supercontinuum generation widely used in many areas including spectroscopy, chemical detection, optical coherent tomography, etc. \cite{OEA-2-2-180020-1,Sylvestre:21}. A crucial component in generating ultrashort pulses is the fiber amplifier, where the main challenge is managing high nonlinear phase shifts without wave breaking while achieving pulse durations of a few hundred femtoseconds or less. The nonlinear evolution of ultrashort pulses in the fiber amplifier requires a comprehensive approach that combines experimental research with theoretical analysis. 

Traditionally, mathematical modeling has been the primary tool for analyzing pulse propagation in optical fibers, playing a vital role in understanding the underlying physical processes. However, as a pulse propagates through an amplifier, its peak power can become high enough to excite various nonlinear effects, the interplay of which leads to complex pulse dynamics, including soliton fission and significant spectral broadening \cite{RevModPhys.78.1135}. Third-order dispersion, Raman scattering, and self-steepening are the primary destabilizing factors. Third-order dispersion breaks the symmetry of the soliton pulse, causing the emission of dispersive waves, leading to energy loss from the soliton and spectral broadening \cite{PhysRevA.79.023824}. Intrapulse Raman scattering induces a spectral redshift among the emerging solitons and drives energy exchange between temporally separated solitons, modifying their widths and frequencies \cite{Tai:88}. Self-steepening distorts the pulse shape, leading to asymmetric evolution and pulse instability \cite{PhysRevA.79.063840}. In addition, environmental perturbations \cite{PhysRevLett.93.183901}, higher-order nonlinear and dispersion \cite{ROY20093798} effects can further complicate soliton dynamics and make spectrum shape more intricate. Despite the fact that the models for pulse propagation in amplifiers are well developed and capable to interpret complex soliton dynamics, it is very difficult to reach high accuracy in reproducing experimental results because the nonlinear coefficients and all experimental conditions are not always known with sufficient precision \cite{RevModPhys.78.1135,Finot:07,Sidorenko:19,TomaszewskaRolla2022,Singh2018}. This limits the predictive capabilities of numerical models and creates a niche for alternative approaches that can achieve closer quantitative agreement with experiments.

Machine learning methods based on artificial neural networks provide a powerful complementary tool for modeling complex systems, particularly when fine-tuned on experimental data. Rather than replacing traditional physical models, deep learning can leverage them, using synthetic data generated from established models to pretrain neural networks. This enables the creation of a ``digital twin'' of a specific experimental setup -- an accurate, data-driven model that captures the system’s behavior even when the physical parameters are not known precisely. %As a result, deep learning models based on multilayer neural networks can potentially provide accurate predictions of high-power pulse dynamics in regimes where traditional numerical models become less accurate or computationally expensive. 

Neural networks have found their primary applications in fields such as image classification, time series analysis, speech synthesis, and others. However, recent years have shown a rapid expansion of the use of neural networks for photonic applications (see \cite{freire2023neural} and references therein). Early studies demonstrated the use of deep learning methods to predict the dynamics of electromagnetic fields in optical fibers, employing unsupervised learning techniques to solve nonlinear Schr\"{o}dinger equations with fully connected neural networks acting as universal function approximators \cite{monterola2001}.
In more recent works, deep learning algorithms have been successfully applied to a variety of photonic problems, including reconstruction of high energy ultrashort laser pulses in frequency-resolved optical gating \cite{stanfield2022}, solving forward and inverse problems of pulse evolution prediction through fiber waveguide \cite{boscolo2021}, predicting key experimental characteristics of supercontinuum generation \cite{salmela2020}, and characterizing radiation features in the modulational instability regime \cite{gautam2021}. 

The main disadvantage of machine learning methods is their requirement for large and diverse training datasets (thousands of samples or more) to ensure good generalization capability. This makes it impractical to train models solely on experimental data, since not all parameters of the experimental system can be continuously varied over a broad range, and certain states of interest may be missed altogether. Previous studies in the area of predicting the nonlinear evolution of the electromagnetic field have relied exclusively on training neural networks with synthetic data generated by numerical simulations \cite{salmela2021predicting,Salmela:22,photonics11020126,pu2023fastpredicting,Sui:23}. However, when models are not validated on actual experimental data, their predictive performance in real-world scenarios remains uncertain. 

To address this limitation, we adopt a transfer learning strategy \cite{Pan:2010} and employ a Transformer decoder architecture for modeling the nonlinear evolution of electromagnetic fields in a real-world fiber amplifier system. Transformers can efficiently process variable-length and even sparse input sequences, making them highly suitable for physical system modeling~\cite{vaswani2017attention}. In the first stage, the Transformer is pretrained on a large synthetic dataset generated through numerical simulations of the amplifier, and in the second stage, it is fine-tuned using experimental measurements. These measurements include spectral  profiles of pulses with varying initial durations and spectral widths, collected from a fiber amplifier whose length is incremented in 10~cm steps. This two-stage approach enables accurate reproduction of experimental results using an experimental dataset that is two orders of magnitude smaller than the synthetic one, achieving high-fidelity reconstruction of both the overall pulse evolution and subtle spectral features. The results demonstrate a significant increase in predictive accuracy for fiber amplifier modeling and highlight the potential of this approach to generalize to other optical systems and experimental configurations.

%\subsection{First Subsection}
%\subsubsection{First Sub Subsection}
%\threesubsection{First lowest-level subsection}

\begin{figure} [t!]
\centering
\includegraphics[width=\linewidth]{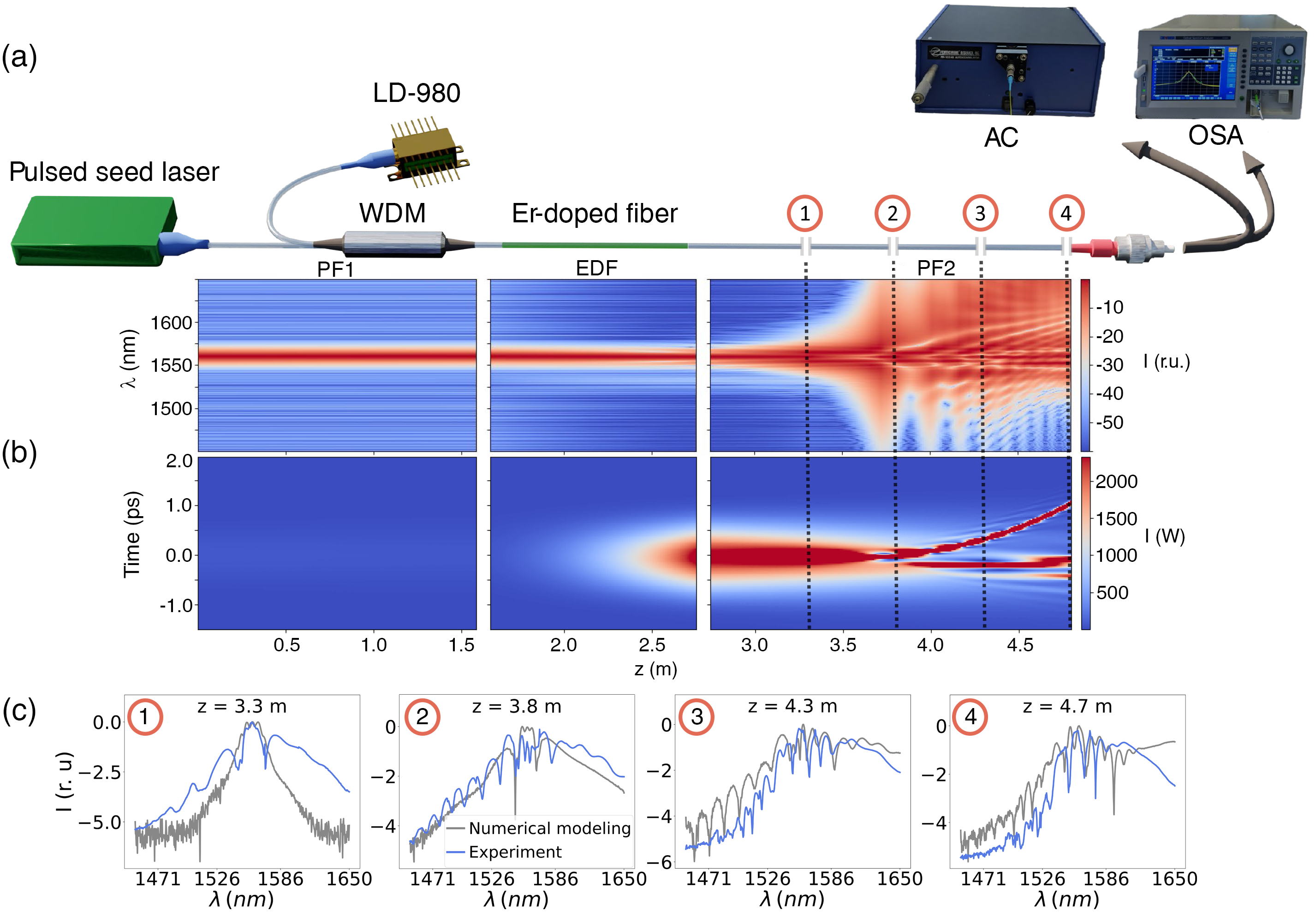}
  \caption{Scheme of the experimental setup. PF1 -- input passive fiber, EDF -- erbium-doped active fiber, PF2 -- output passive fiber, WDM -- optical multiplexer.}
  \label{fig:amplifier}
\end{figure}

\section{Experimental setup}

The experimental setup examined in this work is illustrated in Figure \ref{fig:amplifier}(a). As a seed source we used a self-made carbon nanotube mode-locked Er-doped polarization-maintaining (PM) all-fiber laser oscillator (\cite{GALIAKHMETOVA2021941,Kobtsev:16}), delivering $2\, \text{mW}$ average power pulses at a $3\, \text{MHz}$ repetition rate, corresponding to a pulse energy of $60.6\, \text{pJ}$. A tunable bandpass filter inside the resonator allows varying the duration and peak power of the seed pulse in the range of $0.3$ to $5\, \text{ps}$ and $50$ to $200\, \text{W}$, respectively.

\begin{figure}[h!]
\centering
\includegraphics[width=0.5\textwidth]{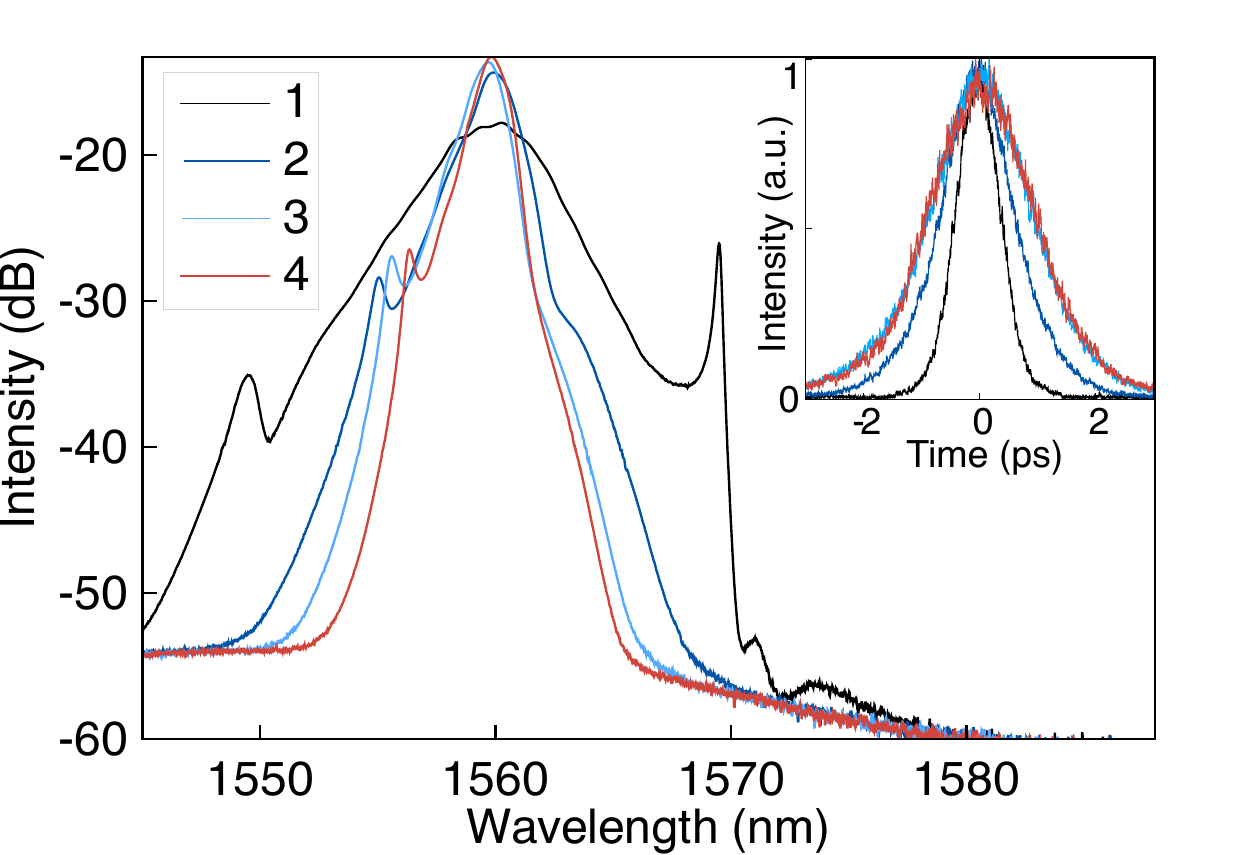}
\caption{Spectral profiles and corresponding autocorrelation traces (insets) of the four pulses (labeled 1-4) at the input of the amplifier.
\label{fig:exp_init_pulses}}
\end{figure}

The seed pulse enters the PM amplifier through $L_{\text{PF1}} = 1.58\,\text{m}$ of the passive fiber and a $980\,\text{nm}$/$1550\,\text{nm}$ wavelength division multiplexer (WDM). It then passes through $L_{\text{EDF}} = 1.17\,\text{m}$ of erbium-doped fiber (EDF) with a core absorption of $80\,\text{dB/m}$, pumped by a $980\,\text{nm}$ diode. To collect the experimental dataset, we varied the output passive fiber length $L_{\text{PF2}}$ from $55\,\text{cm}$ to $205\,\text{cm}$ in $10\,\text{cm}$ increments. For each $L_{\text{PF2}}$, we measured the pulse width and spectrum of four different input seed pulses, each with an energy of $60\,\text{pJ}$ but distinct temporal durations (from $0.8$ to $2\,\text{ps}$) and spectral widths, as shown in Fig.~\ref{fig:exp_init_pulses}. These measurements were repeated for amplifier pump currents ranging from $400$ to $800\,\text{mA}$ in $100\,\text{mA}$ increments, resulting in 300 measured spectral profiles, autocorrelation functions (ACFs), and output power values.
The corresponding pump powers for these currents were 0.195, 0.250, 0.306, 0.359, and 0.414 W, respectively.

An example of the calculated temporal and spectral evolution of a pulse with an initial width of 0.8 ps along the amplifier, obtained using conventional numerical simulations based on the nonlinear Schrödinger equation, is shown in Fig.~\ref{fig:amplifier}(b). The details of the numerical model can be found in the Methods section (subsection 7.1).  As the pulse is amplified in the active fiber, the soliton order~\cite{AGRAWAL2013497} changes from 1.9 to 14.1, leading to an imbalance between dispersion and nonlinearity. After entering the passive fiber section, the nonlinearity leads to pulse narrowing and fission, accompanied by significant broadening of the optical spectrum. The comparison of numerical and experimental spectra in the passive fiber segment is shown in Fig. 1(c). Overall, the spectral evolution shows qualitative agreement; however, even at the shortest propagation distance, noticeable differences between the experimental and numerical spectral shapes are observed. These discrepancies can be attributed to the emergence of higher-order nonlinear effects not accounted for in the simulations, or to inaccuracies in the fiber parameters. The sequence of experimental spectra will be further used to fine-tune the Transformer model for accurate prediction of pulse evolution in the amplifier.

%\begin{figure}[h]
%\centering
%\includegraphics[width=0.6\textwidth]%{figs/fields_spec_temp.png}
%\caption{Evolution of spectral (a) and temporal (b) intensities along the amplifier. $P_0=192$~W, $T_0=0.3$~ps, $P_p=0.414$~W. 
%\label{fig:evol}}
%\end{figure}

% Figure \ref{fig:amplifier} shows the experimental setup of an erbium-doped fiber amplifier (EDF) pumped by a laser diode at a wavelength of $\sim980\,\text{nm}$. The lengths of the input passive fiber and active fiber are $L_{PF1} = 0.47\,\text{m}$ and $L_{EDF} = 1.17\,\text{m}$, respectively. The output passive fiber length, $L_{PF2}$, was varied in the experiment... \textcolor{red}{(to do: Misha)}

\section{Architecture of the Neural Network}

We investigate the potential of a Transformer decoder architecture for the prediction of the nonlinear evolution of the electromagnetic field in the fiber amplifier. Originally developed for natural language processing, the Transformer architecture is built around the self-attention mechanism, which enables the model to analyze each input element in the context of all others. Like LSTMs (long short-term memory networks), Transformers can capture long-range dependencies, but with greater scalability and computational efficiency, thanks to parallelizable computations and improved learning dynamics \cite{vaswani2017attention}. Unlike LSTMs, which require strictly sequential computation, fixed input lengths, and continuity in the data, Transformers handle variable-length and even sparse input sequences with ease, making them well-suited for physical system modeling.

\begin{figure}[h]
\centering
\includegraphics[width=\textwidth]{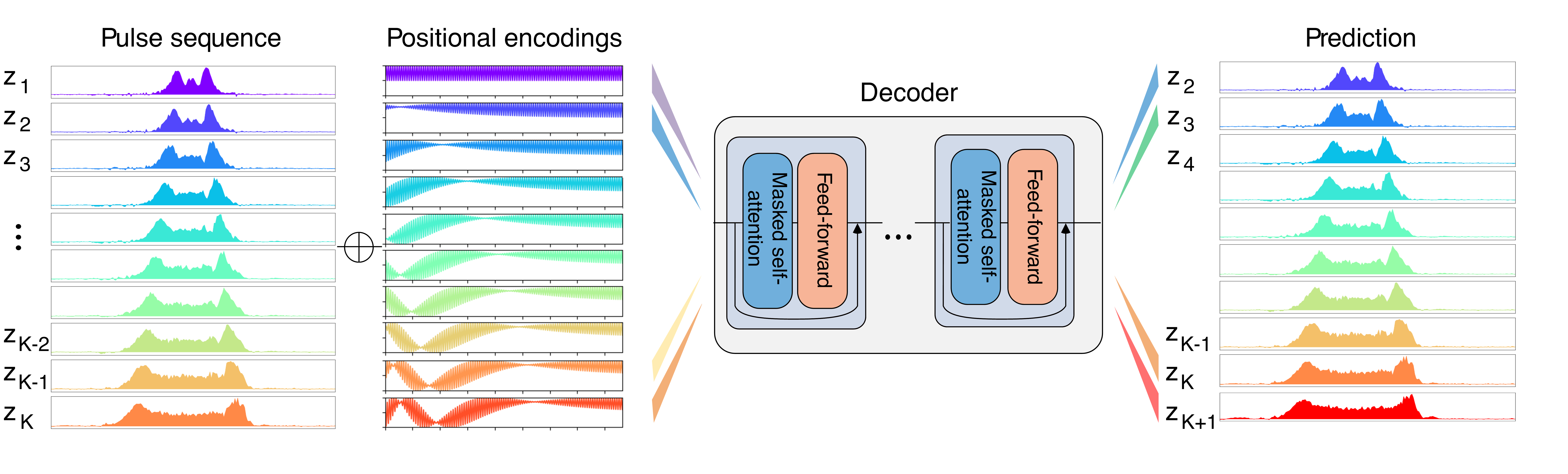}
\caption{Scheme of the neural network architecture with a self-attention mechanism.
\label{fig:NN_scheme}}
\end{figure}

The implemented Transformer-based autoregressive decoder architecture consists of multiple stacked decoder blocks, a standard design for sequence prediction tasks such as word generation. Each block contains two main components: masked self-attention and a fully connected feedforward network (see Fig.~\ref{fig:NN_scheme} and Fig.~\ref{fig:transformer} in the Methods section for further details). This architecture serves as the backbone of the GPT family of large language models \cite{radford2018improving, radford2019language}. The masked self-attention mechanism ensures that, when computing dependencies within the input sequence, the model does not attend to future elements, thereby preserving the causal structure necessary for autoregressive prediction.

The input to the network is a sequence of $K$ spectral profiles measured along the fiber at positions $z_i$, where $i = 1, \ldots, K$. The output is a sequence of the $K$ spectral profiles, shifted forward along the propagation axis: $z_i$, where $i = 2, \ldots, K+1$. This output is then recursively fed back into the model as input. Using this autoregressive strategy, we can predict the evolution of the signal over an arbitrary number of steps.

It is important to note that Transformer architecture itself does not inherently recognize the spatial position of each spectral profile within the input sequence. To address this, we use positional encoding: a vector that encodes the spatial coordinate $z_i$ for each spectral profile. This positional encoding vector is added to the corresponding spectral input vector (see Fig.~\ref{fig:NN_scheme} and Eq.~\ref{eq:encoding} in the Methods section). As a result, the model is able to distinguish spectral profiles from different positions and learn how spectral evolution depends on the position along the fiber.

To quantify the predictive performance, we used the mean squared error (MSE) during training and evaluated the model using normalized root mean squared error (NRMSE). The NRMSE is computed for both the final predicted profile and the entire output map along the fiber to assess accuracy in single-step and long-range autoregressive prediction (see Eqs.~\ref{eq:mse}-\ref{eq:nrmse_map} in the Methods section).

\section{Data Collection and Processing}

\subsection{Preparation of synthetic data for model training}

To train and evaluate the neural network, we generated a synthetic dataset using the described numerical model. The fiber parameters were kept fixed, while the input pulse parameters -- modeled with a hyperbolic secant shape -- were varied: the initial pulse peak power  $P_0$ ranged from 50 to 200 W, and the pulse duration $T_0$ was logarithmically spaced from 0.3 to 5 ps. Amplifier pump power was set to 0.195, 0.250, 0.306, 0.359, and 0.414 W, matching the experimental values.

The network operated on spectral intensity data represented in a logarithmic scale. Since the absolute amplitude of the spectral intensity is typically not critical in experimental settings, all spectra were normalized. This normalization reduces the prediction task to forecasting the spectral shape, rather than its absolute magnitude.

The resulting data were interpolated onto coarser computational grids: 512 points for the temporal and spectral coordinates and 50 points for the evolution coordinate, with 22 points corresponding to the passive fiber segment PF2. 

The training and testing datasets were constructed as follows. For each pair of initial conditions \((P_0, T_0)\), the corresponding pulse evolution along the fiber was assigned either to the training or testing set. From each selected evolution, a short sequence of length \(K+1\) was extracted, starting from a randomly chosen position along the fiber (i.e., a random index). The first \(K\) intensity profiles of the sequence were used as input to the neural network, and the \((K+1)\)-th profile served as the target output. This approach ensured that the training and testing sets were completely disjoint, with no overlapping sequences between them.

Three datasets of increasing size and complexity were prepared, each consisting of sequences of length \( K+1 = 6 \):

\begin{itemize}
\item \textbf{Dataset 1} included 1200 sequences. It was constructed using 20 values of \( P_0 \) (50-200 W), 20 values of \( T_0 \) (0.3-5 ps), and three pump power levels: 0.195, 0.306, and 0.414 W.
\item \textbf{Dataset 2} contained 2000 sequences. It extended Dataset 1 by adding two additional pump powers (0.250 and 0.359 W), while keeping the same sets of \( P_0 \) and \( T_0 \).
\item \textbf{Dataset 3} included 4500 sequences. This dataset was expanded both in the number of initial pulse parameters and pump powers: 30 values of \( P_0 \), 30 values of \( T_0 \), and all five pump powers mentioned above.
\end{itemize}

Each dataset was split into training, validation, and test sets using a 60/20/20 ratio. 

The synthetic datasets are significantly broader than the experimental dataset, covering a wider and denser grid of input pulse parameters.

\subsection{Preparation of experimental data for model fine-tuning}

A total of 300 spectral profiles were measured experimentally, representing 60 separate pulse evolutions along the fiber. Given the limited size and the sequential nature of the experimental dataset, a careful data splitting strategy was required to enable effective evaluation of the model prediction.  

\begin{figure}[h!]
\centering
\includegraphics[width=0.75\textwidth]{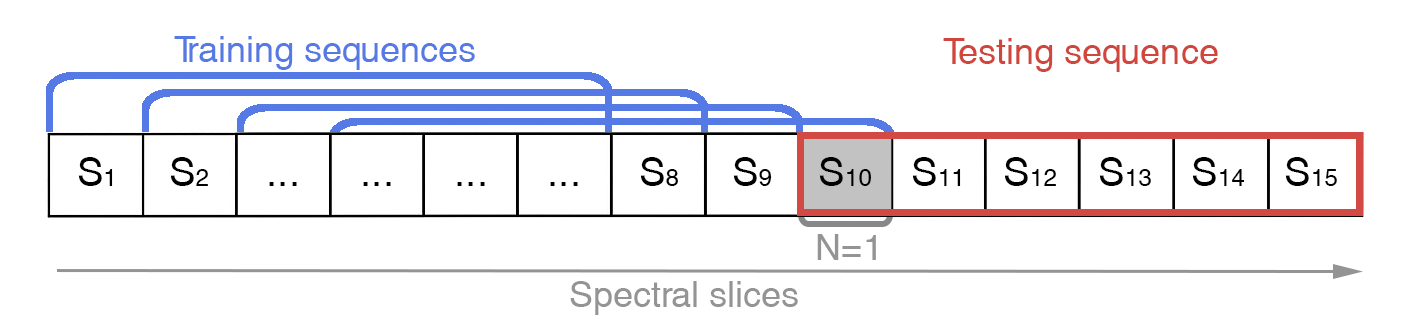}
\caption{Schematic illustration of the experimental data split into training and testing datasets.
\label{fig:exp_dataset_scheme}}
\end{figure}

To construct new datasets from experimental measurements, the following approach was employed (see Fig.~\ref{fig:exp_dataset_scheme}). Each experiment consisted of 15 spectral intensity profiles and was characterized by a specific pulse number (4 pulses) and pump power (5 power levels). For each experiment, one of two splitting strategies was randomly selected: a single sequence of 6 consecutive profiles -- either the first or the last one -- was assigned to the test set. 

The training set was then formed by applying a sliding window of length 6 to the remaining part of the data. Only those sequences that shared no more than \(N\) elements with the test sequence were included. This ensured a controllable degree of overlap between training and testing data, with the parameter \(N\) ranging from 0 (no shared elements) to 5 (maximum overlap of 5 elements).

As a result, each dataset contained 20 test sequences (one per experiment) and \(20(N + 4)\) training sequences of length 6. In each sequence, the first five spectra were used as input to predict the sixth. As the overlap parameter increases, more training sequences are included, leading to improved data efficiency at the cost of greater similarity between training and test sets (see Table~\ref{tab:splitting} in the Methods section).

\section{Results and Discussion}

Training a neural network can be viewed as the process of minimizing a loss function in a high-dimensional parameter space. One of the key challenges in our case is the limited amount of experimental data available: only 20 experiments, each providing 15 measurement points along the fiber. Our central hypothesis is that the minima of the loss function for the synthetic (simulation-based) dataset and the experimental dataset lie in close proximity within the parameter space. Based on this assumption, we adopt a transfer learning approach: the neural network is first trained on a large synthetic dataset generated through numerical modeling, and subsequently fine-tuned using the limited set of experimental measurements. This strategy allows the model to leverage the general physical structure learned from simulations while adapting to real-world conditions with a small amount of experimental data.

\subsection {Training and testing results on synthetic dataset}

To facilitate effective training and help the model converge toward a global minimum of the loss function, we applied several optimization and stabilization techniques. These included the Adam optimizer \cite{kingma2014adam}, hyperparameter tuning \cite{yu2020hyperparameter}, and an inverse square root learning rate schedule with a linear warm-up phase \cite{vaswani2017attention}. Training was carried out on a local server equipped with an NVIDIA RTX 4090 GPU.

\begin{figure}[h]
\centering
\includegraphics[width=0.9\textwidth]{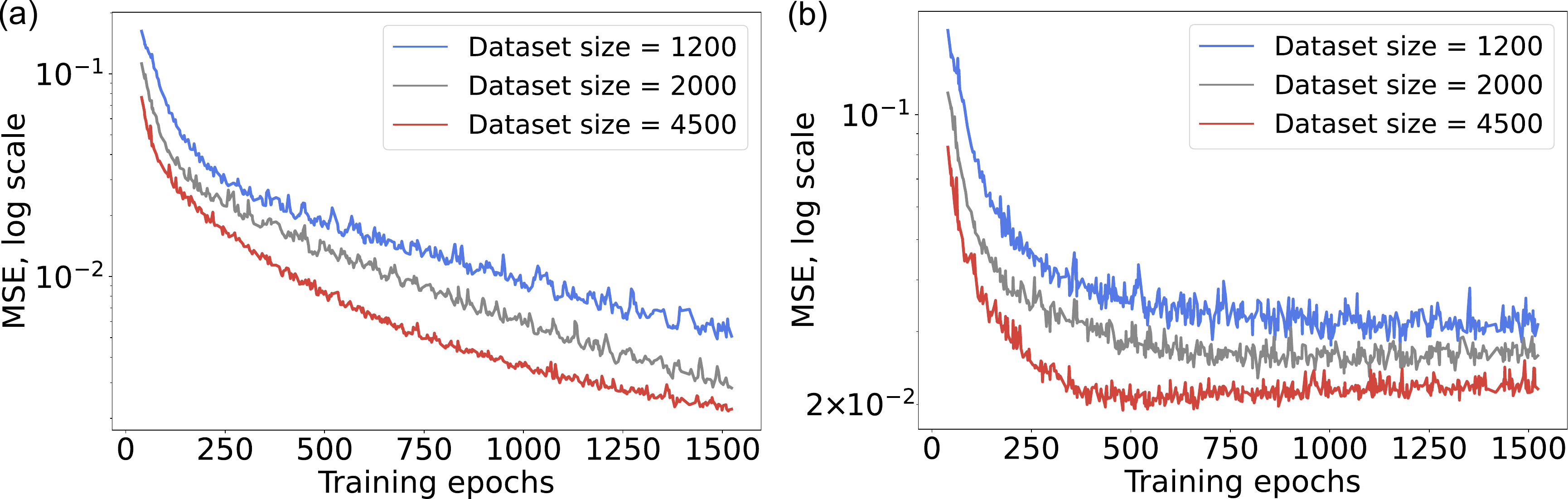}
\caption{Training (a) and validation loss (b) as a function of epochs number on a logarithmic scale.
\label{fig:train_val_loss}}
\end{figure}

Figure~\ref{fig:train_val_loss} illustrates the training and validation loss curves depending on dataset size. In all cases, the training loss steadily decreases, indicating effective model optimization. Models trained on larger datasets achieve lower final loss values, demonstrating improved generalization.  

However, the validation loss curves reveal signs of overfitting after a certain number of epochs, as the validation loss begins to rise. To address overfitting and enhance generalization, we applied early stopping criteria.

The table~\ref{tab:nrmse} presents the integral metrics calculated on test subsets of synthetic data. The results indicate that increasing the size of the synthetic dataset improves prediction accuracy.

\begin{table}[h]
    \centering
    \caption{Metrics of the model calculated on test subsets.}
    \label{tab:nrmse}
    \begin{tabular}{lccc}
        \toprule
        \textbf{Metric} & \multicolumn{3}{c}{\textbf{Dataset size}} \\
        & 1200 & 2000 & 4500 \\
        \midrule
        NRMSE & 2.13e-01 & 2.19e-01 & 2.08e-01 \\
        NRMSE$_{map}$ & 1.33e-01 & 1.34e-01 & 1.22e-01 \\
        \bottomrule
    \end{tabular}
\end{table}

\begin{figure}[h]
\centering
\includegraphics[width=0.9\textwidth]{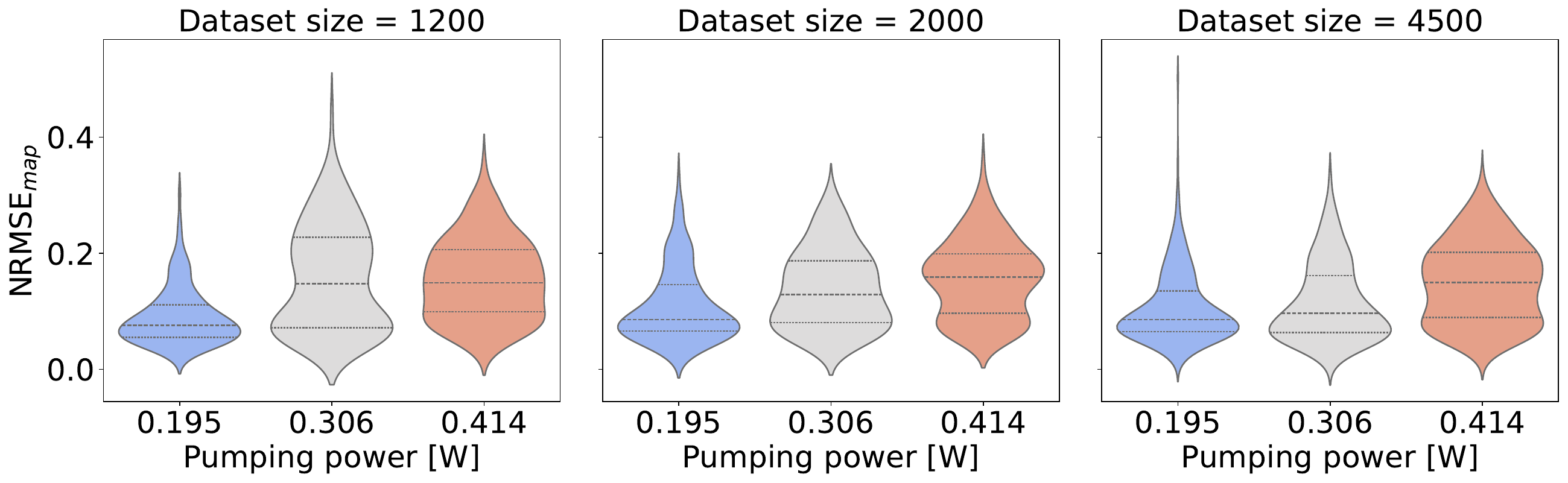}
\caption{Distribution of the normalized root mean squared error (\(\text{NRMSE}_{\text{map}}\)) across different dataset sizes and pumping power levels.
\label{fig:violin}}
\end{figure}

To gain a deeper insight into the distribution of the normalized root mean squared error (\(\text{NRMSE}_{\text{map}}\)) across different dataset sizes and pumping power levels, we generated the violin plots shown in Fig.~\ref{fig:violin}. The results reveal that the error distribution varies with pumping power, suggesting that higher power levels introduce additional complexity in the predictions. Meanwhile, increasing the dataset size reduces the median error and narrows the error spread, indicating enhanced prediction accuracy and stability.

\begin{figure}[h!]
\centering
\includegraphics[width=\textwidth]{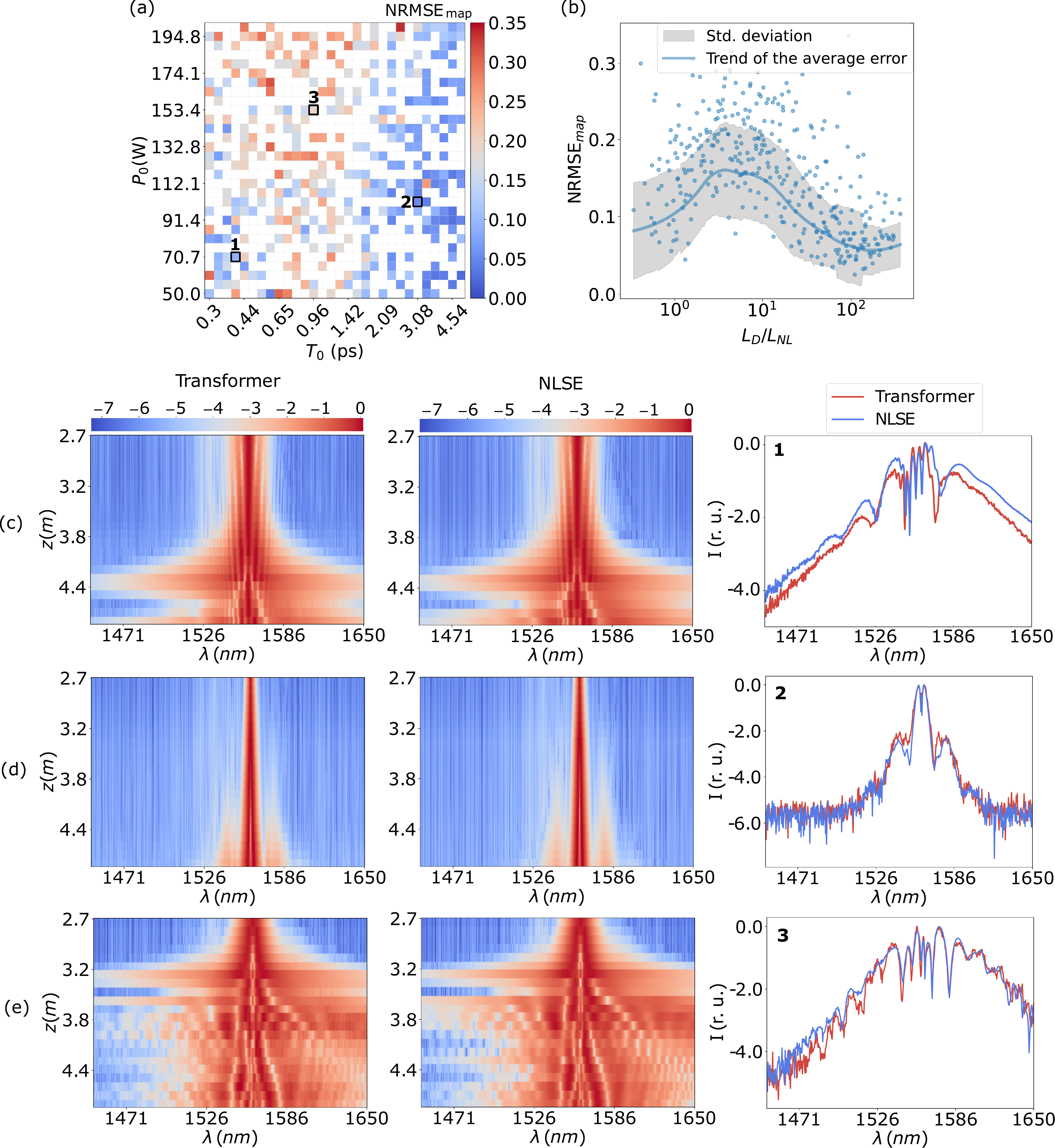}
\caption{Neural network testing results. (a) Error map ($NRMSE_{map}$) in the plain of initial pulse parameters $P_0$-$T_0$ for the test dataset. (b) Distribution of $NRMSE_{map}$ as a function of the ratio $L_D/L_{NL}$. (c-e) Examples of autoregressive predictions for three representative regimes, marked by numbers in panel (a).
\label{fig:error_map}}
\end{figure}

Figure~\ref{fig:error_map}a shows a comprehensive error map in the plane of initial pulse parameters. This map highlights how the initial pulse characteristics influence the overall $NRMSE_{map}$ error. To better interpret this error distribution, it is useful to introduce the dispersion length $L_D=T_0^2/\beta_2$ and the nonlinear length $L_{NL}=1/(\gamma P_0)$ \cite{agrawal2000nonlinear}, which characterize the respective length scales over which dispersive or nonlinear effects become important for pulse evolution. Depending on the relative magnitudes of $L_D$ and $L_{NL}$, pulses can evolve quite differently. When $L_D/L_{NL} \gg 1$, dispersion is negligible and pulse evolution is primarily governed by self-phase modulation (SPM). In this regime, spectral broadening occurs, and modulational instability leads to the formation of two characteristic sidebands. Conversely, when $L_D/L_{NL} \ll 1$, dispersion dominates the dynamics. When $L_D$ and $L_{NL}$ are comparable, both dispersion and nonlinearity interact significantly during propagation. This interplay between group velocity dispersion (GVD) and SPM can result in more complex pulse behavior, including the formation of higher-order solitons and opportunities for pulse compression. As shown in Fig.~\ref{fig:error_map}b, such complex spectral dynamics correspond to regions of higher prediction error.

To compare predicted pulse evolution with the numerical modeling based on NLSE, we picked three distinct regimes of pulse propagation from the data, indicated by the labels "1", "2" and "3" in Fig.~\ref{fig:error_map}a. The corresponding spectral evolution along the fiber, as well as the output spectra for each regime, are presented in Figs.~\ref{fig:error_map}(c)–(e). We see that Transformer can accurately predict spectrum evolution even for complex spectra with a large broadening. However, as previously discussed, achieving agreement with the experimental spectra requires a final step of fine-tuning using the experimental data.

\subsection {Model fine-tuning on experimental dataset}

Fine-tuning on experimental data was performed using the best model previously trained on a synthetic dataset of 4500 samples. To preserve the optimal solution found during pretraining, we initialized fine-tuning with a small learning rate of $10^{-5}$ and applied a linear learning rate scheduler with warm-up. This approach ensured a smooth adaptation to the experimental data without deviating significantly from the pretrained model's parameters.

The impact of fine-tuning was evaluated for different levels of overlap $N$ between training and test sequences (see paragraph 7.5 in the Methods for details). Importantly, even in the strictest case with no overlap, fine-tuning resulted in a substantial 24.5\% reduction in NRMSE compared to the pretrained model, highlighting the robustness of the approach.

\begin{figure}[h!]
\centering
\includegraphics[width=1\textwidth]{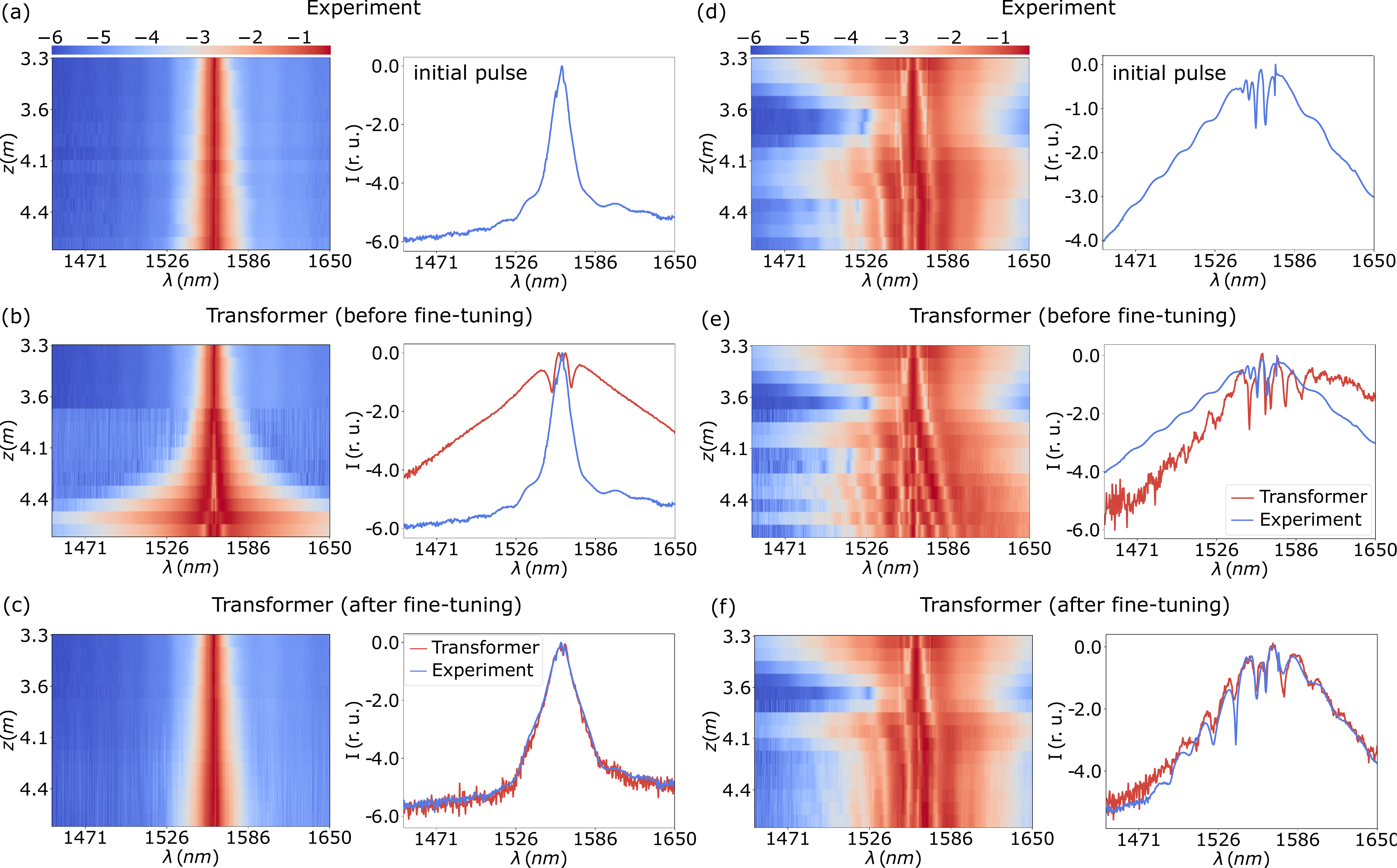}
\caption{Comparison of autoregressive predictions for two representative pulse propagation regimes, before (b,e) and after (c,f) model fine-tuning. Initial pulse shape is shown at the passive fiber $PF_2$ input.
\label{fig:exp_error_map}}
\end{figure}

Figure~\ref{fig:exp_error_map} compares pulse evolution predictions before and after fine-tuning for two distinct propagation regimes under a strict $N=0$ condition (no training-testing overlap). Importantly, the input sequences used for the autoregressive prediction were taken directly from the test set and were never seen by the model during training. 

In each example, the figure compares the ground truth spectral evolution map with the model's predictions before and after fine-tuning. It can be seen that the fine-tuned model more accurately captures both the general structure and fine details of the pulse evolution, significantly reducing prediction error across the propagation length. These results further confirm the model's ability to generalize effectively to unseen experimental data, even in a strictly disjoint training–testing scenario.

\begin{figure}[h!]
\centering
\includegraphics[width=\textwidth]{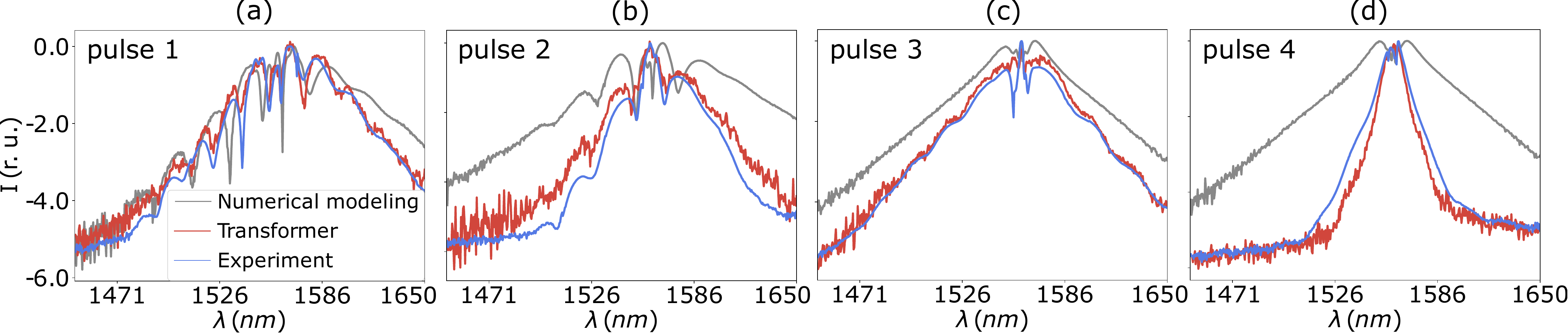}
\caption{Comparison of output spectra for four pulses considered in the experiment: numerical simulation (gray curve), NN prediction (red curve) and experiment (blue curve). Pump power $P_p=306$ mW.
\label{fig:exp_modelling_comparison}}
\end{figure}

Further validating these findings, Fig.~\ref{fig:exp_modelling_comparison} presents output spectra comparisons for all four experimental pulses (pump power $P_p = 306$ mW), contrasting numerical simulations (gray), neural network predictions (red), and experimental measurements (blue).

\begin{figure}[h!]
\centering
\includegraphics[width=\textwidth]{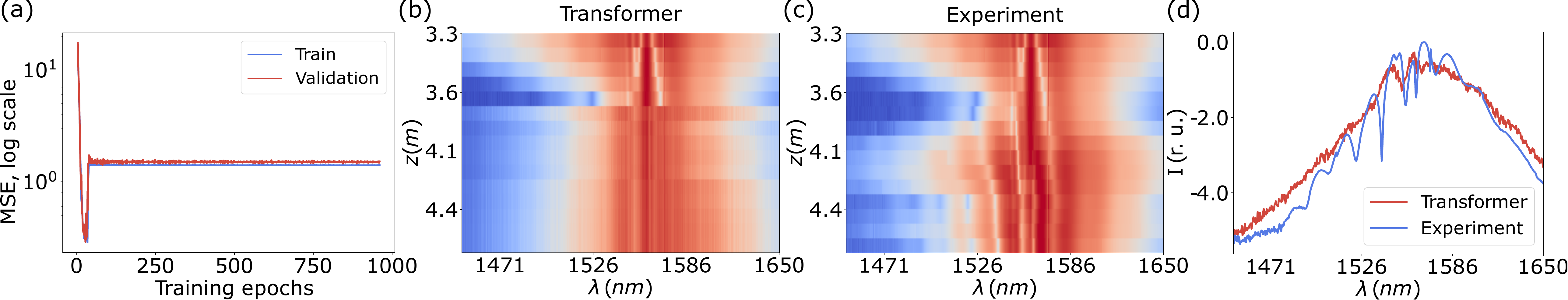}
\caption{Training from scratch on experimental data. (a) Training and validation loss curves as a function of the number of epochs, shown on a logarithmic scale. (b) Example of predicted pulse evolution corresponding to the lowest loss value achieved during training.
\label{fig:exp_scratch_error_map}}
\end{figure}

We also attempted to train the model from scratch using only experimental data. The training curve shown in Fig.~\ref{fig:exp_scratch_error_map}(a) demonstrates the absence of convergence in the loss function, indicating that the model fails to learn a meaningful representation from the limited dataset. Figure~\ref{fig:exp_scratch_error_map}(b) shows an example of the predicted pulse evolution corresponding to the minimum loss value achieved during training (as shown in Fig.~\ref{fig:exp_scratch_error_map}a). As can be seen, the model cannot reproduce the fine spectral structure of the pulse, highlighting the necessity of pretraining on a larger synthetic dataset to ensure adequate generalization.  

\section{Conclusion}

This work demonstrates the effectiveness of a Transformer-based neural network model, trained via transfer learning, for predicting the nonlinear evolution of optical pulses in Er-doped fiber amplifiers under conditions of limited experimental data. By leveraging synthetic data for pretraining and subsequently fine-tuning on a small set of experimental measurements, the proposed approach achieves high prediction accuracy and reliably reconstructs both the overall structure and fine spectral features of pulse evolution. 

Beyond fiber amplifier modeling, the proposed Transformer-based approach holds significant potential for a broad range of ultrafast optics applications that involve complex nonlinear pulse evolution. These include pulse formation dynamics in mode-locked fiber lasers, where real-time spectral measurements can be obtained using dispersive Fourier transform (DFT) techniques, as well as supercontinuum generation in nonlinear fibers and the design of optical frequency combs in both fiber-based and microresonator platforms. In all these cases, the model can serve as a fast and accurate surrogate for solving the forward problem --predicting pulse evolution with given system parameters. More importantly, it offers a promising direction for solving the inverse problem, where one seeks to identify optimal system parameters to achieve a desired output (e.g., maximizing spectral bandwidth or generating specific comb structures). In such settings, where brute-force parameter scanning is typically required, our model can be used to rapidly explore parameter spaces and guide optimization, offering a practical and scalable tool for the design of advanced photonic systems.

% Experimental section

\section{Methods}

\subsection{Experimental characterization}
The characterization of the pulsed generation was performed with the following equipment: Deviser E8600 OSA, ThorLabs InGaAs biased detector DET08CFC and Tektronix MDO3102 $1\, \text{GHz}$ oscilloscope, Femtochrome FR-103WS autocorrelator, ThorLabs S155C photodiode power sensor.

\subsection{Numerical Model of the Amplifier}

To describe pulse propagation along the fiber amplifier, we have performed numerical simulations based on nonlinear Schr\"{o}dinger equations, taking into account the effects of dispersion and nonlinearity:

\begin{equation}
\label{eq:nlse}
\frac{\partial A_{s}(z,t)}{\partial z} =  -i\frac{\beta_2}{2} \frac{\partial^2 A_s(z,t)}{\partial t^2} + i\gamma|A_s(z,t)|^2 A_s(z,t) + \frac{g_s(z)*\tilde{h}(\omega)}{2}A_s(z,t), 
\end{equation}
\noindent{}where $A_s(z,t)$ is the slowly varying envelope associated with the signal, $\beta_2$ is the group velocity dispersion, $\gamma$ is the Kerr nonlinearity, $g_s(z)$ is the signal gain coefficient, and $\tilde{h}(\omega)$ is the lineshape corresponding to the typical gain spectrum in Er-doped fiber. Eq.~(\ref{eq:nlse}) was numerically solved by the split-step Fourier method. The propagation step along the fiber was set to 2~mm. The temporal window considered in the model was equal to 100~ps discretized into $2^{14}$ time steps.

Signal amplification was incorporated using a simplified gain model, which can be derived from the equation for the upper-level population in the steady-state regime and the power balance equation for the propagating signal, followed by a redefinition of the coefficients \cite{Barnard:94}: 

\begin{equation}
\label{eq:amplification_simple}
g_s(z) = \frac{{g_0 \cdot P_p / {P_p}^{\text{sat}} - \alpha}}{{1 + P_s(z) / {P_s}^{\text{sat}} + P_p / {P_p}^{\text{sat}}}} - \alpha^{*}.
\end{equation}

Here, $g_0$ is the small-signal gain coefficient, which was measured to be approximately $25\, \text{dB}$ at a wavelength of $1560\, \text{nm}$. The saturable absorption coefficient $\alpha$ was found to be around $42.5\,\text{dB}$ at $1560\,\text{nm}$, while the coefficient of non-saturable losses $\alpha^{*}$ can be neglected in this case. All fiber parameters are listed in Table~\ref{tab:experimental}.

\begin{table}[h]
\centering
\begin{tabular}{|l|l|l|}
\hline
\textbf{Fiber} & \textbf{Parameter} & \textbf{Value} \\
\hline
\multirow{3}{*}{Active fiber (EDF)} 
& Dispersion (${\beta_{2}}_{EDF}$) &  $-22.4\, \text{ps}^2/\text{km}$ \\
& Nonlinearity ($\gamma_{EDF}$) &  $1.95\, \text{W}^{-1}\text{km}^{-1}$ \\
& Pump saturation power (${P_{p}}^{\text{sat}}$) &  $31.7\, \text{mW}$ \\
& Signal saturation power (${P_{s}}^{\text{sat}}$) & $3.4\, \text{mW}$ \\
& Small signal gain ($g_0$) & $40\, \text{dB}$ \\
& Saturable absorption ($alpha$) & $42.5\, \text{dB}$ \\
\hline
\multirow{2}{*}{Passive fiber (PF2)} 
& Dispersion (${\beta_{2}}_{PF}$) &  $-24.5\, \text{ps}^2/\text{km}$ \\
& Nonlinearity ($\gamma_{PF}$) & $1.7\, \text{W}^{-1}\text{km}^{-1}$ \\
\hline
\end{tabular}
\caption{Parameters of active and passive fibers}
\label{tab:experimental}
\end{table}

The input to the amplifier was a pulse with a hyperbolic secant shape. To ensure that the spectral width of the initial pulse in the simulations matched the experimentally measured spectral width, a chirp was added to the pulse phase by propagating it through a dispersive fiber of 5 meters in length with a dispersion of \(\beta_2 = -20\) ps\(^2\)/km.  

We optimized the parameters of the amplification model (Eq.~(\ref{eq:amplification_simple})) and performed numerical simulations of pulse propagation in the amplifier. Although the simulations qualitatively agree with the experimental observations, quantitative agreement was not achieved. The discrepancy between the simulated and experimental spectra increases with pump power and also depends on the parameters of the input pulse.

\subsection{Architecture of the Neural Network}

\begin{figure}[h!]
\centering
\includegraphics[width=0.7\textwidth]{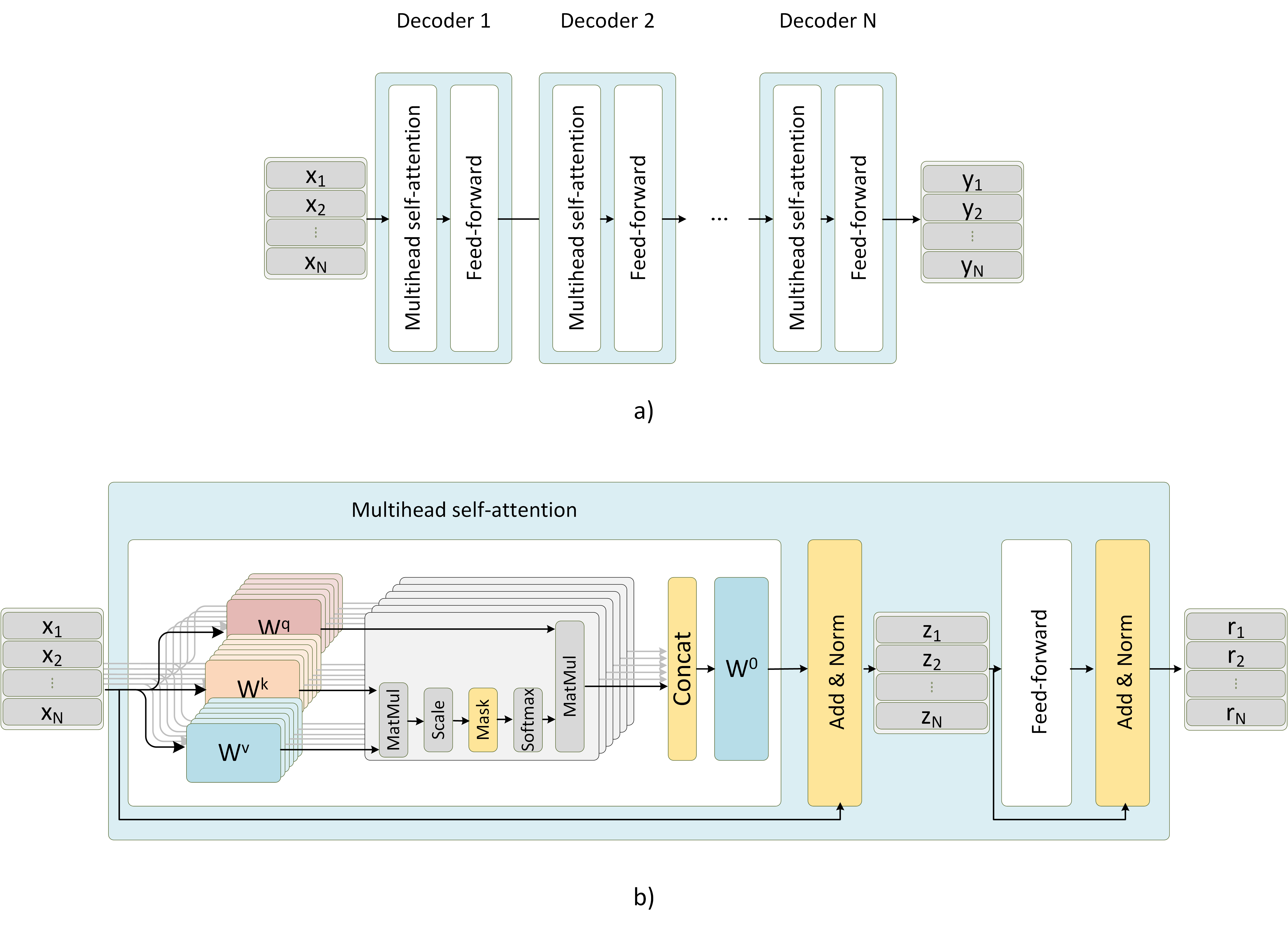}
\caption{Transformer decoder architecture diagram: (a) the overall model structure consisting of blocks with a self-attention mechanism, (b) the structure of an individual self-attention block. 
\label{fig:transformer}}
\end{figure}

In our implementation, we used a Transformer decoder consisting of six decoder blocks, each with 8-dimensional masked self-attention (Fig.~\ref{fig:transformer}). To provide a theoretical foundation, we begin by outlining the principle of self-attention, where each element of the input sequence is represented as a vector. The self-attention mechanism operates using three different representations of the input sequence: queries (Q), keys (K), and values (V). During computation, the current focus of the attention mechanism is treated as a query, while the surrounding context vectors serve as keys. The result is computed based on the corresponding values.

\vspace{-3pt}

Let $X$ be the matrix of input vectors with the shape $n \times d_{\text{model}}$, where $n$ is the number of vectors and $d_{\text{model}}$ is the dimensionality of each vector. The input vectors are projected into query, key, and value representations using learned weight matrices $W^Q$, $W^K$, and $W^V$, each of size $d_{\text{model}} \times d_k$, where $d_k$ is the dimensionality of the internal representation:

$$
Q = X \cdot W^Q, \quad K = X \cdot W^K, \quad V = X \cdot W^V.
$$

The influence of context vectors on the current position is determined by the dot product between the query and each key vector. These similarity scores are used to compute attention weights, which are then applied to the corresponding value vectors:

\begin{equation}
\label{eq:attention}
\text{Attention}(Q, K, V) = \text{softmax}\left(\frac{QK^T}{\sqrt{d_k}}\right)V.
\end{equation}

Here, the softmax function transforms the similarity scores into probabilities:

$$
\text{softmax}(x_i) = \frac{\exp(x_i)}{\sum_{j} \exp(x_j)}.
$$

The output of the attention layer is then passed through a feedforward layer to restore the original dimensionality of the input.

The multi-head self-attention mechanism extends this concept by allowing the model to attend to information from multiple representation subspaces simultaneously. This is achieved by applying several sets of weight matrices $W^Q$, $W^K$, and $W^V$ in parallel (Fig.\ref{fig:transformer}(b)):

$$
\text{MultiHead}(Q, K, V) = [\text{head}_1, \text{head}_2, ..., \text{head}_h] W^O,
$$

$$
\text{head}_i = \text{Attention}(XW^Q_i, XW^K_i, XW^V_i).
$$

Here, $h$ is the number of attention heads, and $W^O$ is a projection matrix that combines the concatenated outputs of all heads into a single output vector.

To preserve the causal structure of the sequence, masking is applied to prevent the model from attending to future positions when computing attention. Specifically, in Eq.~\ref{eq:attention}, the matrix $QK^T$ is modified before applying softmax by masking out elements above the main diagonal using an upper-triangular matrix:

\begin{equation}
{QK^T}_{ij} = \begin{cases}
-\infty, & \text{if } i < j \\
q_i \cdot k_j, & \text{otherwise}.
\end{cases}
\label{eq:mask}
\end{equation}

This ensures that the attention mechanism only uses information from the current and previous positions, which is essential for autoregressive prediction.

In addition, absolute positional encoding was incorporated to inject information about the relative position of tokens within the input sequence. As proposed in the original Transformer architecture \cite{vaswani2017attention}, the encoding is defined as:

\begin{equation}
\text{PE}(\text{pos}, i) = 
\begin{cases}
\sin\left(\frac{\text{pos}}{10000^{2j/d_{\text{model}}}}\right), & \text{if } i = 2j, \\
\cos\left(\frac{\text{pos}}{10000^{2j/d_{\text{model}}}}\right), & \text{if } i = 2j+1,
\end{cases}
\label{eq:encoding}
\end{equation}

where \(\text{PE}\) is the positional encoding vector of dimension \( d_{\text{model}} \), \(\text{pos}\) is the position of an element in the sequence, and \( i \) is the feature index of the sequence element (ranging from 0 to \( d_{\text{model}} - 1 \)). The sine function is applied to even-indexed dimensions, while the cosine function is used for odd-indexed dimensions. This approach allows the model to incorporate information about the relative positions of tokens in a sequence without relying on recurrence or convolution.

\subsection{Metrics used for tracking NN performance}

Here, we introduce the metrics used to evaluate the final predictive performance of the trained neural network. The network is designed to predict a single pulse profile by capturing the evolution dynamics from a sequence of profiles, optimizing its predictions using the mean squared error (MSE) loss:

\begin{equation}
\label{eq:mse}
MSE(I, \hat{I}) = \frac{1}{N}\sum_{i=1}^{N_{\text{domain}}} (I_{i} - \hat{I}_{i})^{2},
\end{equation}

where \( I \) and \( \hat{I} \) represent the predicted and true spectral intensity arrays at a fixed \( z \)-coordinate, respectively, and \( N_{\text{domain}} \) denotes the spectral domain size.

To assess the accuracy of the model in autoregressive prediction, we use the normalized root mean squared error (NRMSE) metric. The error is calculated using two approaches: for individual intensity profiles at the amplifier output and by averaging over all positions along the passive fiber ($PF_2$):

\begin{equation}
\label{eq:nrmse}
NRMSE(I, \hat{I}) = \sqrt{\frac{ \sum\limits_{i=1}^{N_{\text{domain}}} (I_i - \hat{I}_i)^2}
                    {\sum\limits_{i=1}^{N_{\text{domain}}} \hat{I}_i^2}}, 
\end{equation}

\begin{equation}
\label{eq:nrmse_map}
NRMSE_{\text{map}}(I, \hat{I}) = \sqrt{\frac{ \sum\limits_{i=1}^{N_{\text{domain}}} \sum\limits_{j=1}^{N_{\text{length}}} (I_{ij} - \hat{I}_{ij})^2}
                    {\sum\limits_{i=1}^{N_{\text{domain}}} \sum\limits_{j=1}^{N_{\text{length}}} \hat{I}_{ij}^2}}, 
\end{equation}

where \( I \) and \( \hat{I} \) represent the predicted and true spectral intensity arrays at a fixed \( z \)-coordinate, respectively. Here, \( N_{\text{domain}} \) denotes the spectral domain size and \( N_{\text{length}} \) represents the spatial domain size.

\subsection{Results of fine-tuning with varying overlap between test and train}

Table~\ref{tab:train_test_ratio} summarizes the resulting train-test ratio for different values of $N$. As the overlap parameter increases, more training sequences are included, leading to improved data efficiency at the cost of greater similarity between training and testing sets.

\begin{table}[h]
\centering
\caption{Train-test ratio depending on the number of overlapping elements $N$}
\label{tab:train_test_ratio}
    \begin{tabular}{cccc}
    \toprule
    \textbf{Number of overlapping elements $N$} & \textbf{Train} & \textbf{Test} & \textbf{Train : Test Ratio} \\
    \midrule
    0 & 80  & 20 & 4 : 1 \\
    1 & 100 & 20 & 5 : 1 \\
    2 & 120 & 20 & 6 : 1 \\
    3 & 140 & 20 & 7 : 1 \\
    4 & 160 & 20 & 8 : 1 \\
    5 & 180 & 20 & 9 : 1 \\
    \bottomrule
    \end{tabular}
\label{tab:splitting}
\end{table}

Table~\ref{tab:metrics_comparison_tuned_model} summarizes the normalized root mean square error on experimental test sequences for models fine-tuned with varying $N$. As the number of overlapping elements increased from 0 to 5 (corresponding to 0\% to 83.3\% overlap), prediction accuracy improved consistently. 

\begin{table}[h]
\centering
\caption{NRMSE on test sequences from experimental data for fine-tuned models}
\label{tab:metrics_comparison_tuned_model}
    \begin{tabular}{lc}
    \toprule
    \textbf{Overlap percentage (number of overlapping elements $N$)} & \textbf{NRMSE} \\
    \midrule
    Before fine-tuning & 11.2e-02 \\  
    \midrule
    0\% (0)   & 8.45e-02 \\
    16.7\% (1) & 8.20e-02 \\
    33.3\% (2) & 5.51e-02 \\
    50\% (3)   & 4.45e-02 \\
    66.7\% (4) & 4.05e-02 \\
    83.3\% (5) & 3.89e-02 \\
    \bottomrule
    \end{tabular}
\end{table}

\begin{figure}[h!]
\centering
\includegraphics[width=0.4\textwidth]{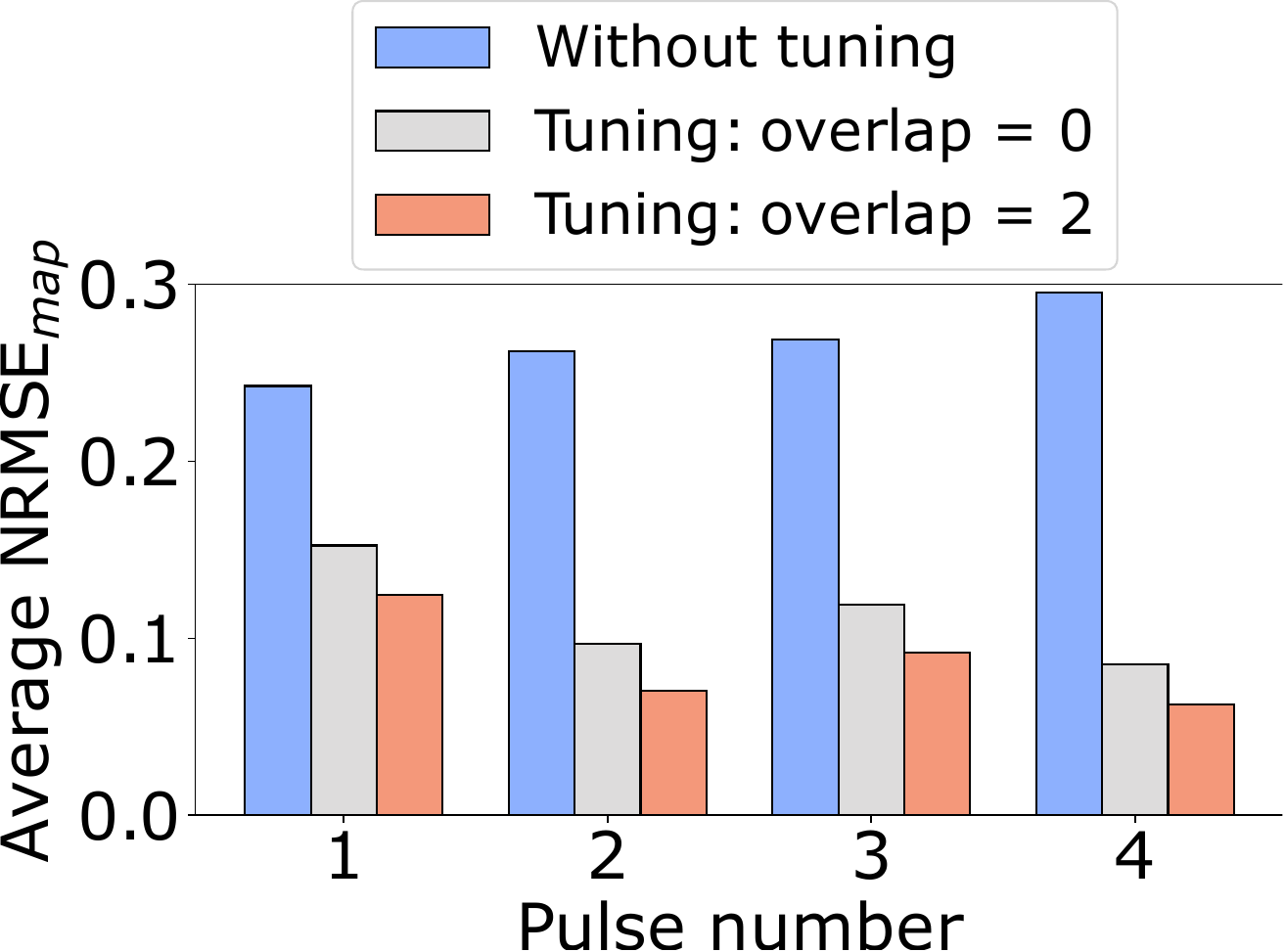}
\caption{Comparison of average $NRMSE_{map}$ for each of the four experimental pulses, before and after model fine-tuning.
\label{fig:nrmse_comparison}}
\end{figure}

To further illustrate this effect, Figure~\ref{fig:nrmse_comparison} shows the average $NRMSE_{map}$ computed over the full spectral evolution maps for each of the four experimental pulses. The figure compares the pretrained model with two fine-tuned models corresponding to overlap levels $N=0$ and $N=2$. As seen in the figure, fine-tuning significantly enhances accuracy across all pulse types, with the best performance achieved at moderate overlap. This confirms that even limited access to nearby data points during training can improve generalization without leading to excessive data leakage.

%\threesubsection{Second part of experimental section}\\

\medskip
\textbf{Supporting Information} \par %Please delete the Suppporting Information statement if it is not applicable. Please supply Supporting Information in another file. Supporting information should not be provided in .tex format
Supporting Information is available from the Wiley Online Library or from the author.

\medskip
\textbf{Author contributions} \par
Conceptualization, A.B. and Y.G.; methodology, A.B., A.G., K.S. and A.R.; investigation, M.M., A.G. and A.M.; software, K.S. and A.G.; validation, A.B., A.G. and K.S.; data curation, A.G., M.M. and A.B., visualization, A.G., M.M., A.B. and K.S., writing--original draft preparation, A.B., M.M, Y.G., A.R., writing – review \& editing, A.R., M.M., Y.G., A.N., A.M., A.B., resources, A.R., Y.G. and A.N. All authors have read and agreed to the published version of the manuscript.

% Acknowledgements
\medskip
\textbf{Acknowledgements} \par %delete if not applicable))
A.G. and A.B. acknowledge the support of the Russian Science Foundation (№ 25-61-00010, https://rscf.ru/project/25-61-00010/). A.R. and K.S. acknowledge the support of the Ministry of Science and Higher Education of the Russian Federation (Project No. FSUS-2021-0015).

\medskip
\textbf{Conflict of Interest} \par
The authors declare no conflict of interest.

\medskip
\textbf{Data Availability Statement} \par
The data that support the findings of this study are openly available in Figshare at  \\https://doi.org/10.6084/m9.figshare.c.7955465.v5, reference number 7955465.

% References
\medskip

% Use the following code if you wish to generate your bibliography with BibTeX;
% replace the string "MSP-template" below with the name(s) of
% the BibTeX data base(s) you want to use.
% The resulting bibliography-output (the content of the .bbl file)
% must be pasted back into this file before submission.
% Please also include your BibTeX data base file(s) in your submission
% so that we can re-run BibTeX if necessary.
%
\bibliographystyle{MSP}
\bibliography{bibliography}

@article{richardson2010,
author = {D. J. Richardson and J. Nilsson and W. A. Clarkson},
journal = {J. Opt. Soc. Am. B},
number = {11},
pages = {B63--B92},
publisher = {Optica Publishing Group},
title = {High power fiber lasers: current status and future perspectives \[Invited\]},
volume = {27},
month = {Nov},
year = {2010},
url = {https://opg.optica.org/josab/abstract.cfm?URI=josab-27-11-B63},
doi = {10.1364/JOSAB.27.000B63},
}

@article{fu2018,
author = {Walter Fu and Logan G. Wright and Pavel Sidorenko and Sterling Backus and Frank W. Wise},
journal = {Opt. Express},
number = {8},
pages = {9432--9463},
publisher = {Optica Publishing Group},
title = {Several new directions for ultrafast fiber lasers \[Invited\]},
volume = {26},
month = {Apr},
year = {2018},
url = {https://opg.optica.org/oe/abstract.cfm?URI=oe-26-8-9432},
doi = {10.1364/OE.26.009432},
}

@article{freire2023neural,
author = {Pedro Freire and Egor Manuylovich and Jaroslaw E. Prilepsky and Sergei K. Turitsyn},
journal = {Adv. Opt. Photon.},
number = {3},
pages = {739--834},
publisher = {Optica Publishing Group},
title = {Artificial neural networks for photonic applications---from algorithms to implementation: tutorial},
volume = {15},
month = {Sep},
year = {2023},
doi = {10.1364/AOP.484119},
}

@article{monterola2001,
author = {Christopher Monterola and Caesar Saloma},
journal = {Opt. Express},
number = {2},
pages = {72--84},
publisher = {Optica Publishing Group},
title = {Solving the nonlinear Schrodinger equation with an unsupervised neural network},
volume = {9},
month = {Jul},
year = {2001},
url = {https://opg.optica.org/oe/abstract.cfm?URI=oe-9-2-72},
doi = {10.1364/OE.9.000072},
}

@article{stanfield2022,
author = {Matthew Stanfield and Jordan Ott and Christopher Gardner and Nicholas F. Beier and Deano M. Farinella and Christopher A. Mancuso and Pierre Baldi and Franklin Dollar},
journal = {Scientific Reports},
number = {1},
title = {Real-time reconstruction of high energy, ultrafast laser pulses using deep learning},
volume = {12},
article-number = {5299},
year = {2022},
url = {https://doi.org/10.1038/s41598-022-09041-y},
doi = {10.1038/s41598-022-09041-y},
}

@article{boscolo2021,
title = {Modelling self-similar parabolic pulses in optical fibres with a neural network},
journal = {Results in Optics},
volume = {3},
pages = {100066},
year = {2021},
issn = {2666-9501},
doi = {https://doi.org/10.1016/j.rio.2021.100066},
url = {https://www.sciencedirect.com/science/article/pii/S2666950121000146},
author = {Sonia Boscolo and John M. Dudley and Christophe Finot},
}

@article{salmela2020,
author = {Lauri Salmela and Coraline Lapre and John M. Dudley and Goëry Genty},
journal = {Scientific Reports},
number = {1},
title = {Machine learning analysis of rogue solitons in supercontinuum generation},
volume = {10},
article-number = {9596},
year = {2020},
url = {https://doi.org/10.1038/s41598-020-66308-y},
doi = {10.1038/s41598-020-66308-y},
}

@article{gautam2021,
title = {Comparative study of neural network architectures for modelling nonlinear optical pulse propagation},
journal = {Optical Fiber Technology},
volume = {64},
pages = {102540},
year = {2021},
issn = {1068-5200},
doi = {https://doi.org/10.1016/j.yofte.2021.102540},
url = {https://www.sciencedirect.com/science/article/pii/S1068520021000894},
author = {Naveenta Gautam and Amol Choudhary and Brejesh Lall},
}

@article{vaswani2017attention,
  title={Attention is all you need},
  author={Vaswani, Ashish and Shazeer, Noam and Parmar, Niki and Uszkoreit, Jakob and Jones, Llion and Gomez, Aidan N and Kaiser, {\L}ukasz and Polosukhin, Illia},
  journal={Advances in neural information processing systems},
  volume={30},
  year={2017}
}

@article{radford2018improving,
  title={Improving language understanding by generative pre-training},
  author={Radford, Alec and Narasimhan, Karthik and Salimans, Tim and Sutskever, Ilya and others},
  year={2018},
  publisher={OpenAI}
}

@article{radford2019language,
  title={Language models are unsupervised multitask learners},
  author={Radford, Alec and Wu, Jeffrey and Child, Rewon and Luan, David and Amodei, Dario and Sutskever, Ilya and others},
  journal={OpenAI blog},
  volume={1},
  number={8},
  pages={9},
  year={2019}
}

@article{salmela2021predicting,
  title={Predicting ultrafast nonlinear dynamics in fibre optics with a recurrent neural network},
  author={Salmela, Lauri and Tsipinakis, Nikolaos and Foi, Alessandro and Billet, Cyril and Dudley, John M and Genty, Go{\"e}ry},
  journal={Nature machine intelligence},
  volume={3},
  number={4},
  pages={344--354},
  year={2021},
  publisher={Nature Publishing Group UK London}
}

@article{photonics11020126,
AUTHOR = {Saraeva, Karina and Bednyakova, Anastasia},
TITLE = {Enhanced bi-LSTM for Modeling Nonlinear Amplification Dynamics of Ultra-Short Optical Pulses},
JOURNAL = {Photonics},
VOLUME = {11},
YEAR = {2024},
NUMBER = {2},
ARTICLE-NUMBER = {126},
URL = {https://www.mdpi.com/2304-6732/11/2/126},
ISSN = {2304-6732},
DOI = {10.3390/photonics11020126}
}

@incollection{agrawal2000nonlinear,
  title={Nonlinear fiber optics},
  author={Agrawal, Govind P},
  booktitle={Nonlinear Science at the Dawn of the 21st Century},
  pages={195--211},
  year={2000},
  publisher={Springer}
}

@article{pu2023fastpredicting,
author = {Pu, Guoqing and Liu, Runmin and Yang, Hang and Xu, Yongxin and Hu, Weisheng and Hu, Minglie and Yi, Lilin},
title = {Fast Predicting the Complex Nonlinear Dynamics of Mode-Locked Fiber Laser by a Recurrent Neural Network with Prior Information Feeding},
journal = {Laser \& Photonics Reviews},
volume = {17},
number = {6},
pages = {2200363},
doi = {https://doi.org/10.1002/lpor.202200363},
url = {https://onlinelibrary.wiley.com/doi/abs/10.1002/lpor.202200363},
eprint = {https://onlinelibrary.wiley.com/doi/pdf/10.1002/lpor.202200363},
year = {2023}
}

@article{GALIAKHMETOVA2021941,
title = {Direct measurement of carbon nanotube temperature between fiber ferrules as a universal tool for saturable absorber stability investigation},
journal = {Carbon},
volume = {184},
pages = {941-948},
year = {2021},
issn = {0008-6223},
doi = {https://doi.org/10.1016/j.carbon.2021.08.032},
url = {https://www.sciencedirect.com/science/article/pii/S0008622321008241},
author = {Diana Galiakhmetova and Yuriy Gladush and Aram Mkrtchyan and Fedor S. Fedorov and Eldar M. Khabushev and Dmitry V. Krasnikov and Raghavan Chinnambedu-Murugesan and Egor Manuylovich and Vladislav Dvoyrin and Alex Rozhin and Mark Rümmeli and Sergey Alyatkin and Pavlos Lagoudakis and Albert G. Nasibulin}
}

@article{Kobtsev:16,
author = {Sergey Kobtsev and Aleksey Ivanenko and Yury G. Gladush and Boris Nyushkov and Alexey Kokhanovskiy and Anton S. Anisimov and Albert G. Nasibulin},
journal = {Opt. Express},
number = {25},
pages = {28768--28773},
publisher = {Optica Publishing Group},
title = {Ultrafast all-fibre laser mode-locked by polymer-free carbon nanotube film},
volume = {24},
month = {Dec},
year = {2016},
url = {https://opg.optica.org/oe/abstract.cfm?URI=oe-24-25-28768},
doi = {10.1364/OE.24.028768},
}

@ARTICLE{Barnard:94,
  author={Barnard, C. and Myslinski, P. and Chrostowski, J. and Kavehrad, M.},
  journal={IEEE Journal of Quantum Electronics}, 
  title={Analytical model for rare-earth-doped fiber amplifiers and lasers}, 
  year={1994},
  volume={30},
  number={8},
  pages={1817-1830},
  doi={10.1109/3.301646}}

@article{Sui:23,
author = {Hao Sui and Hongna Zhu and Huanyu Jia and Qi Li and Mingyu Ou and Bin Luo and Xihua Zou and Lianshan Yan},
journal = {Opt. Lett.},
number = {18},
pages = {4889--4892},
publisher = {Optica Publishing Group},
title = {Predicting nonlinear multi-pulse propagation in optical fibers via a lightweight convolutional neural network},
volume = {48},
month = {Sep},
year = {2023},
url = {https://opg.optica.org/ol/abstract.cfm?URI=ol-48-18-4889},
doi = {10.1364/OL.496973},
}

@ARTICLE{Pan:2010,
  author={Pan, Sinno Jialin and Yang, Qiang},
  journal={IEEE Transactions on Knowledge and Data Engineering}, 
  title={A Survey on Transfer Learning}, 
  year={2010},
  volume={22},
  number={10},
  pages={1345-1359},
  doi={10.1109/TKDE.2009.191}}

@article{Salmela:22,
author = {Lauri Salmela and Mathilde Hary and Mehdi Mabed and Alessandro Foi and John M. Dudley and Go\"{e}ry Genty},
journal = {Opt. Lett.},
number = {4},
pages = {802--805},
publisher = {Optica Publishing Group},
title = {Feed-forward neural network as nonlinear dynamics integrator for supercontinuum generation},
volume = {47},
month = {Feb},
year = {2022},
url = {https://opg.optica.org/ol/abstract.cfm?URI=ol-47-4-802},
doi = {10.1364/OL.448571},
}

@article{RevModPhys.78.1135,
  title = {Supercontinuum generation in photonic crystal fiber},
  author = {Dudley, John M. and Genty, Go\"ery and Coen, St\'ephane},
  journal = {Rev. Mod. Phys.},
  volume = {78},
  issue = {4},
  pages = {1135--1184},
  numpages = {0},
  year = {2006},
  month = {Oct},
  publisher = {American Physical Society},
  doi = {10.1103/RevModPhys.78.1135},
  url = {https://link.aps.org/doi/10.1103/RevModPhys.78.1135}
}

@article{PhysRevA.79.023824,
  title = {Dispersive waves emitted by solitons perturbed by third-order dispersion inside optical fibers},
  author = {Roy, Samudra and Bhadra, S. K. and Agrawal, Govind P.},
  journal = {Phys. Rev. A},
  volume = {79},
  issue = {2},
  pages = {023824},
  numpages = {6},
  year = {2009},
  month = {Feb},
  publisher = {American Physical Society},
  doi = {10.1103/PhysRevA.79.023824},
  url = {https://link.aps.org/doi/10.1103/PhysRevA.79.023824}
}

@article{Tai:88,
author = {Kuochou Tai and Akira Hasegawa and Naoaki Bekki},
journal = {Opt. Lett.},
number = {5},
pages = {392--394},
publisher = {Optica Publishing Group},
title = {Fission of optical solitons induced by stimulated Raman effect},
volume = {13},
month = {May},
year = {1988},
url = {https://opg.optica.org/ol/abstract.cfm?URI=ol-13-5-392},
doi = {10.1364/OL.13.000392},
}

@article{PhysRevA.79.063840,
  title = {Wave-turbulence approach of supercontinuum generation: Influence of self-steepening and higher-order dispersion},
  author = {Barviau, Benoit and Kibler, Bertrand and Picozzi, Antonio},
  journal = {Phys. Rev. A},
  volume = {79},
  issue = {6},
  pages = {063840},
  numpages = {12},
  year = {2009},
  month = {Jun},
  publisher = {American Physical Society},
  doi = {10.1103/PhysRevA.79.063840},
  url = {https://link.aps.org/doi/10.1103/PhysRevA.79.063840}
}

@article{PhysRevLett.93.183901,
  title = {Vector Soliton Fission},
  author = {Lu, F. and Lin, Q. and Knox, W. H. and Agrawal, Govind P.},
  journal = {Phys. Rev. Lett.},
  volume = {93},
  issue = {18},
  pages = {183901},
  numpages = {4},
  year = {2004},
  month = {Oct},
  publisher = {American Physical Society},
  doi = {10.1103/PhysRevLett.93.183901},
  url = {https://link.aps.org/doi/10.1103/PhysRevLett.93.183901}
}

@article{ROY20093798,
title = {Perturbation of higher-order solitons by fourth-order dispersion in optical fibers},
journal = {Optics Communications},
volume = {282},
number = {18},
pages = {3798-3803},
year = {2009},
issn = {0030-4018},
doi = {https://doi.org/10.1016/j.optcom.2009.06.018},
url = {https://www.sciencedirect.com/science/article/pii/S0030401809005677},
author = {Samudra Roy and Shyamal K. Bhadra and Govind P. Agrawal},
}

@article{Finot:07,
author = {Christophe Finot and Benoit Barviau and Guy Millot and Alexej Guryanov and Alexej Sysoliatin and Stefan Wabnitz},
journal = {Opt. Express},
number = {24},
pages = {15824--15835},
publisher = {Optica Publishing Group},
title = {Parabolic pulse generation with active or passive dispersion decreasing optical fibers},
volume = {15},
month = {Nov},
year = {2007},
url = {https://opg.optica.org/oe/abstract.cfm?URI=oe-15-24-15824},
doi = {10.1364/OE.15.015824},
}

@article{Sidorenko:19,
author = {Pavel Sidorenko and Walter Fu and Frank Wise},
journal = {Optica},
number = {10},
pages = {1328--1333},
publisher = {Optica Publishing Group},
title = {Nonlinear ultrafast fiber amplifiers beyond the gain-narrowing limit},
volume = {6},
month = {Oct},
year = {2019},
url = {https://opg.optica.org/optica/abstract.cfm?URI=optica-6-10-1328},
doi = {10.1364/OPTICA.6.001328},
}

@article{ORAZI2021543,
title = {Ultrafast laser manufacturing: from physics to industrial applications},
journal = {CIRP Annals},
volume = {70},
number = {2},
pages = {543-566},
year = {2021},
issn = {0007-8506},
doi = {https://doi.org/10.1016/j.cirp.2021.05.007},
url = {https://www.sciencedirect.com/science/article/pii/S0007850621001232},
author = {L. Orazi and L. Romoli and M. Schmidt and L. Li},
}

@Article{OEA-2-2-180020-1,
title = {Recent development of flat supercontinuum generation in specialty optical fibers},
journal = {Opto-Electronic Advances},
volume = {2},
number = {2},
pages = {180020-1-180020-9},
year = {2019},
issn = {2096-4579},
doi = {10.29026/oea.2019.180020},	
url = {https://www.oejournal.org/oea/en/article/doi/10.29026/oea.2019.180020},
author = {Huanhuan Liu and Ye Yu and Wei Song and Qiao Jiang and Fufei Pang}
}

@article{Sylvestre:21,
author = {T. Sylvestre and E. Genier and A. N. Ghosh and P. Bowen and G. Genty and J. Troles and A. Mussot and A. C. Peacock and M. Klimczak and A. M. Heidt and J. C. Travers and O. Bang and J. M. Dudley},
journal = {J. Opt. Soc. Am. B},
number = {12},
pages = {F90--F103},
publisher = {Optica Publishing Group},
title = {Recent advances in supercontinuum generation in specialty optical fibers \[Invited\]},
volume = {38},
month = {Dec},
year = {2021},
url = {https://opg.optica.org/josab/abstract.cfm?URI=josab-38-12-F90},
doi = {10.1364/JOSAB.439330},
}

@incollection{AGRAWAL2013497,
title = {Chapter 12 - Novel Nonlinear Phenomena},
booktitle = {Nonlinear Fiber Optics (Fifth Edition)},
publisher = {Academic Press},
pages = {497-552},
year = {2013},
issn = {15575837},
doi = {https://doi.org/10.1016/B978-0-12-397023-7.00012-7},
author = {Govind Agrawal}
}

@article{Singh2018,
  author    = {Singh, Neetesh and Xin, Ming and Vermeulen, Diedrik R. G. and Shtyrkova, Katia and Li, Nanxi and Callahan, Patrick T. and Magden, Emir Salih and Ruocco, Alfonso and Fahrenkopf, Nicholas and Baiocco, Christopher and Kuo, Bill P.-P. and Radic, Stojan and Ippen, Erich P. and K{\"a}rtner, Franz X. and Watts, Michael R.},
  title     = {Octave-spanning coherent supercontinuum generation in silicon on insulator from 1.06 μm to beyond 2.4 μm},
  journal   = {Light: Science \& Applications},
  year      = {2018},
  volume    = {7},
  pages     = {17131},
  doi       = {10.1038/lsa.2017.131}
}

@article{TomaszewskaRolla2022,
  author    = {Dorota Tomaszewska-Rolla and Robert Lindberg and Valdas Pasiskevicius and Fredrik Laurell and Grzegorz Soboń},
  title     = {A comparative study of an {Yb}-doped fiber gain-managed nonlinear amplifier seeded by femtosecond fiber lasers},
  journal   = {Scientific Reports},
  year      = {2022},
  volume    = {12},
  number    = {1},
  pages     = {404},
  doi       = {10.1038/s41598-021-04420-3},
  url       = {https://doi.org/10.1038/s41598-021-04420-3},
  publisher = {Nature Publishing Group},
  issn      = {2045-2322}
}

@article{kingma2014adam,
  title={Adam: A Method for Stochastic Optimization},
  author={Kingma, Diederik P. and Ba, Jimmy},
  journal={arXiv preprint arXiv:1412.6980},
  year={2014}
}

@article{yu2020hyperparameter,
  title={Hyper-Parameter Optimization: A Review of Algorithms and Applications},
  author={Tong Yu and Hong Zhu},
  journal={arXiv preprint arXiv:2003.05689},
  year={2020},
  url={https://arxiv.org/abs/2003.05689}
}

\end{document}